\documentstyle[11pt,newpasp,twoside,epsf]{article}
\def\edcomment#1{\iffalse\marginpar{\raggedright\sl#1\/}\else\relax\fi}
\reversemarginpar

\begin{document}
\title{Constraints in the $\lambda_{0}$-$\Omega_{0}$ plane from 
gravitational lensing}
\author{Phillip Helbig}
\affil{Rijksuniversiteit Groningen, Kapteyn Instituut, Postbus 800,
NL-9700 AV Groningen, The Netherlands}

\begin{abstract}
I review simultaneous constraints on the cosmological parameters
$\lambda_{0}$ and $\Omega_{0}$ from gravitational lensing.  The emphasis
is on systematic extragalactic surveys for strong gravitational lenses,
mainly the largest and best-defined such survey, JVAS/CLASS. 
\end{abstract}

\section{Introduction}

Since the details of the gravitational-lensing effect depend on the
cosmological model, it offers a means of determining the cosmological
parameters $H_{0}$, $\lambda_{0}$ and $\Omega_{0}$ by comparing
expectations from different cosmological models with observations.
Advantages of using gravitational lensing to learn about the
cosmological model include the fact that it is based on relatively well
understood astrophysics and that it makes use of information from an
intermediate redshift range, complementing tests which use information
primarily from the low-redshift (e.g.\ cosmic flows) or high-redshift
(e.g.\ cosmic microwave background fluctuations) regimes. 

I review simultaneous constraints on the cosmological parameters
$\lambda_{0}$ and $\Omega_{0}$ from gravitational lensing.  (Constraints
on $H_{0}$ are discussed in other contributions in this volume; in
particular, Schechter discusses constraints from lensing.)  The emphasis
is on systematic extragalactic surveys for strong gravitational lenses,
mainly the largest and best-defined such survey, JVAS/CLASS. However,
other methods of constraining $\lambda_{0}$ using (mostly strong)
gravitational lensing are also discussed.  After briefly reviewing the
basic theory and history of the subject, I present the currently
available constraints and briefly (since this is covered in other
contributions in this volume, such as that by Lineweaver) touch on joint
constraints with other cosmological tests. Finally, I discuss systematic
errors and prospects for the future. 

In general, one can derive constraints on $\lambda_{0}$ and $\Omega_{0}$
from gravitational lensing when there is more than one source plane
involved (e.g. Golse, Kneib, \& Soucail 2000).  This is the case in most
examples of weak lensing and cluster lensing.  Also, arc statistics
(e.g.\ Bartelmann et al. 1998) can give information about the
cosmological model, not only since more than one source plane is
involved, but also since this is sensitive to the evolution of structure
(the lenses in this context), which also depends on the cosmological
model.  Another possibility is provided by higher-order effects in
measuring $H_{0}$ through gravitational-lens time delays; the time delay
is inversely proportional to $H_{0}$, but there is a weaker, non-linear
dependence on $\lambda_{0}$ and $\Omega_{0}$.  Finally, the statistical
analysis of gravitational-lens surveys provides a potentially powerful
method of measuring the cosmological parameters.

\section{Time Delay}

The basic idea behind using the time delay in a gravitational-lens
system, i.e.\ the time between seeing variations in the brightness of
one image and seeing similar variations in the brightness of another
image, is a simple one:  Most observables in a gravitational-lens system
(angles, brightness ratios etc) are dimensionless; measuring the time
delay provides the scale for the lens system.  Cosmological distances
depend linearly on (the reciprocal of) $H_{0}$\ but $\lambda_{0}$\ and
$\Omega_{0}$\ enter at higher order.  Thus, measuring time delays in
systems with various values of $z_{\rm d}$ and $z_{\rm s}$ might enable
one to put constraints on $\lambda_{0}$ and $\Omega_{0}$.  This
technique was already mentioned by Refsdal (1966) but at present there
are too few lens systems and too many observational and lens-modelling
uncertainties for this method to provide useful constraints on
$\lambda_{0}$ and $\Omega_{0}$.  For a discussion of future prospects, 
see Haarsma, Leh\'{a}r, \& Barkana (2000).

The basic equation is
\begin{equation}
H_{0} = (\Delta t)^{-1}Tf \qquad,
\end{equation}
where $f$ is a quantity which depends only on observables and the lens
model and $T$ is the cosmological correction function (Refsdal 1966,
Kayser \& Refsdal 1983).  Since 
\begin{equation}
T = \frac{H_{0}}{c}
\frac{D_{\rm d}D_{\rm s}}{D_{\rm ds}}
(1+z_{d})
\frac{z_{\rm s}-z_{\rm d}}{z_{\rm d}z_{\rm s}} \qquad,
\end{equation}
there is the natural behaviour
\begin{displaymath}
T\rightarrow 0 \quad \mbox{for} \quad z_{\rm s} \rightarrow 0 \qquad.
\end{displaymath}

\section{Gravitational-Lensing Statistics}

\subsection{Basic Principles}

Gravitational-lensing statistics---the number of gravitational lenses
found in a survey and their properties such as image configuration
(including multiplicity, image separations and flux ratios), source and
lens redshifts, nature of lens galaxies etc---depends on the
cosmological model, the properties of the lens galaxy population, the
properties of the source population and on selection effects.  The
influence of the cosmological is quite large, since two effects occur
which tend to reinforce each other (see, e.g., Appendix A in Quast \&
Helbig (1999)): 
\begin{itemize} 
\item 
The volume element $\frac{\mathrm{d}V}{\mathrm{d}z}$ influences the
number of lenses.  It is of course the redshift $z$ and related
quantities which are observed.  The volume element as a function of
redshift is strongly dependent on the cosmological model.  Thus, if one
fixes the space density of galaxies at $z=0$, varying the cosmological
parameters can greatly vary the number of potential lens galaxies per
redshift interval at higher redshift.  (Of course, if one has {\em
observed\/} the number of potential lens galaxies per redshift interval
at higher redshift, then this is an additional constraint; here, I
assume that the space density of galaxies is fixed at $z=0$ and free,
i.e.\ determined by the cosmological model through
$\frac{\mathrm{d}V}{\mathrm{d}z}$, at higher redshift.) 
\item
The cross section of an individual galaxy depends on a combination of
various angular size distances.  The dependence of the angular size
distances on the cosmological model (e.g. Kayser, Helbig, \& Schramm
1997) thus means that the lensing cross section of an individual galaxy
has a dependence on the cosmological model. 
\end{itemize}
The total lensing cross section obviously depends on the number of
potential lens galaxies and the cross section of an individual galaxy.
Both of these depend on the cosmological model and the two effects tend
to reinforce each other. 

In general, the larger $\lambda_{0}$\, the more lenses one expects,
especially for $\lambda_{0} > 0$. The number of lenses, however, is only
one measurable quantity and due to parameter degeneracy, just `counting
lenses' will give weaker constraints. Traditionally, lensing statistics
has tended to give comparatively tight upper limits on $\lambda_{0}$,
especially in the often studied flat-universe case.  It is important to
keep in mind that, as with other constraints in the
$\lambda_{0}$-$\Omega_{0}$ plane, the actual values for the best-fit
model are less interesting than more robust limits (cf. Helbig 1999). 

Following the formalism in Kochanek (1996), for each source in a survey
one can calculate the probability $p_{\mathrm{lens}}$ that it is a lens
system with the observed properties; for non-lens-systems, the
probability that they are non-lenses is obviously $1 -
p_{\mathrm{lens}}$.  For a given cosmological model, the total
likelihood is just the product of the likelihoods for the individual
objects in the survey, i.e.\ information from both the lenses and the
non-lenses is used (see, e.g., Appendix A in Quast \& Helbig (1999)).
One can thus assign a relative probability to each cosmological model.
Of course, there is a dependence on quantities other than $\lambda_{0}$
and $\Omega_{0}$ as well.  In the new results presented below, we
consider all other variables to take fixed, observationally determined
values and concentrate on the corresponding constraints in the
$\lambda_{0}$-$\Omega_{0}$ plane (see Helbig et al.\ (1999) and
references therein for more details).

\subsection{Some History}

Turner, Ostriker, \& Gott (1984) presented the first quantitative
lensing-statistics analysis, but assumed $\lambda_{0}=0$.  Fukugita et
al.\ (1992) generalised the Turner et al.\ treatment to non-zero
$\lambda_{0}$, and concluded that $\lambda_{0} < 0.95$.  Kochanek (1996)
came to the more quantitative conclusions that $\lambda_{0} < 0.66$
(95\%) for $k = 0 $ and $\Omega_{0} > 0.15$ (90\%) for $\lambda_{0} =
0$. This paper is in some sense the definitive analysis, but it is
important to remember that it is not based on the best input data.  In
particular, it is based on optical quasar surveys.  Not only is the
$m$-$z$ relation for QSOs not particularly well-known---especially since
the $m$-$z$ relation should apply to QSOs with the same selection
criteria as those in the gravitational-lens survey---but selection
effects in optical gravitational-lens surveys are more difficult to
quantify.  It should be noted that Falco, Kochanek, \& Mu\~noz (1998)
obtain a higher value of $\lambda_{0}$ based on radio data.  This gives
some idea of the uncertainties involved.  Using only radio data, Falco
et al.\ (1998) obtain $\lambda_{0} < 0.73$ ($2\sigma$) for $k = 0$;
using combined radio and optical data, they get $\lambda_{0} < 0.62$
($2\sigma$) for $k = 0$. This paper is also interesting since it
quantitatively details how the constraints change depending on various
assumptions.  Helbig et al.\ (1999) presented an analysis based on the
JVAS gravitational-lens survey, which is reasonably large but also well
understood.  Since in the interesting part of parameter space, lensing
statistics essentially measures $\lambda_{0} - \Omega_{0}$, reducing the
results to just a few numbers gives $-2.69 < \lambda_{0} - \Omega_{0} <
0.68$ (95\%); for $k = 0$: $-0.85 < \lambda_{0} < 0.84$. 
\begin{figure}[t]
\plotone{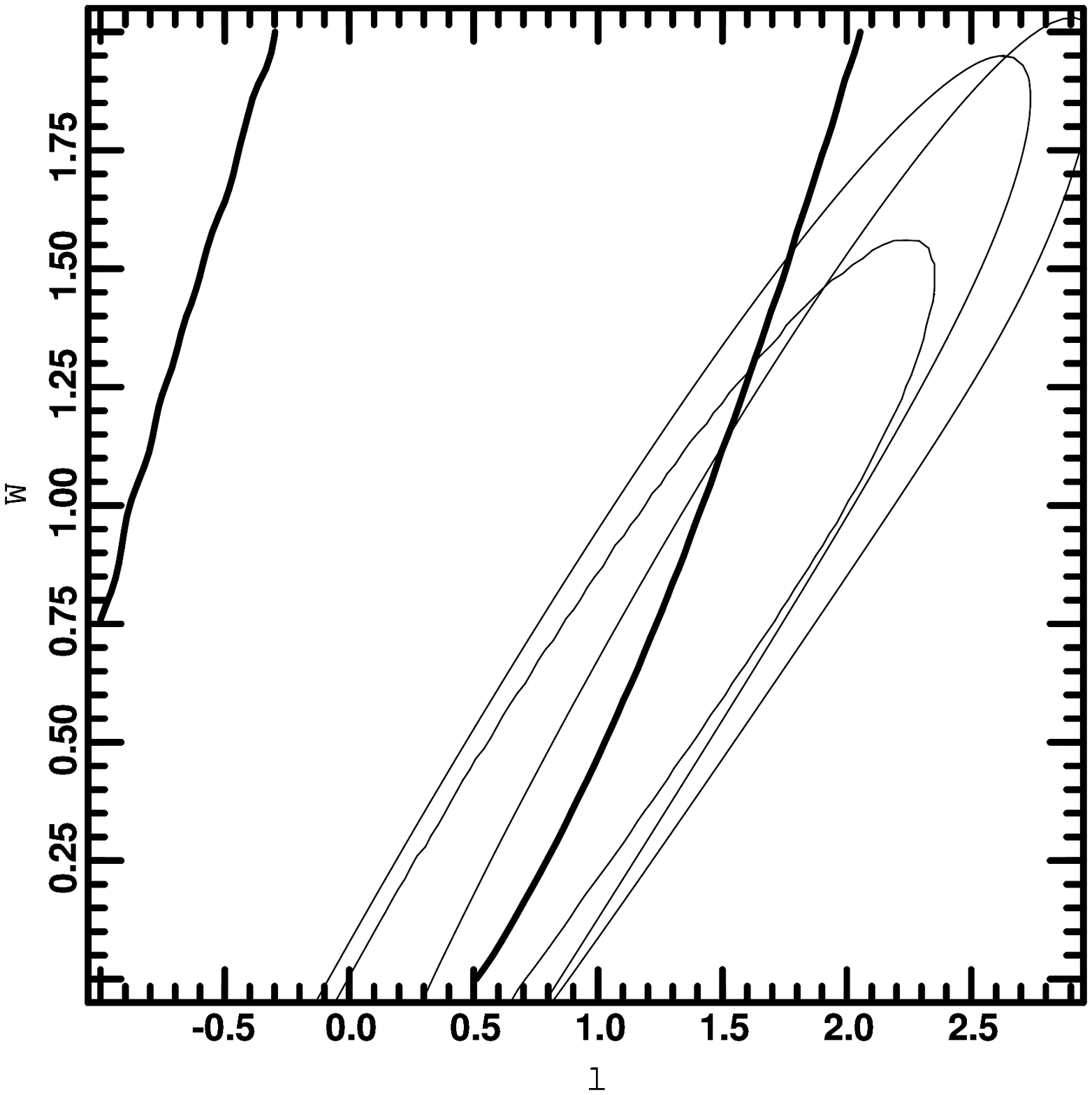}
\caption{Degeneracy of constraints in the $\lambda_{0}$-$\Omega_{0}$
plane for gravitational-lensing statistics (thick curves) and the
$m$-$z$ relation for type Ia supernovae (three different results).  All
curves are 95\% confidence contours.  See Helbig (1999), from which this
figure is taken, for more details.} 
\end{figure}

In passing, it should be remembered that the {\em redshift of lens
galaxies alone\/} is also sensitive to the cosmological model, but the
situation here is not clear (see Helbig 2000a for discussion and
references).

\subsection{Radio Lens Surveys}

There are many good reasons to do a gravitational-lens survey in the
radio as opposed to optical: 
\begin{itemize}
\item Using interferometry, the beam size is much smaller than the image
separation. 
\item Flat-spectrum objects are compact (i.e. (almost) point sources),
allowing typical lensing morphologies to be recognised easily. 
\item Somewhat related to the previous point, the lensing probability
depends on the cross section determined by the lens population; for
extended sources, the source geometry partially determines what is
recognised as a lens system. 
\item Most sources are quasars at high redshift, which leads to a high
lensing rate. 
\item Since flat-spectrum objects are compact, they can be variable on
relatively short timescales, which aids in determining time delays. 
\item There is no bias from lens galaxies due to extinction by the lens
or comparable brightness of source and lens, as can be the case with
optical surveys. 
\item High-resolution followup is possible with interferometers such as
MERLIN, the VLBA, VLBI\dots. 
\end{itemize}
But there is one disadvantage:
\begin{itemize}
\item Additional work is required to get redshifts.
\end{itemize}

By far the largest gravitational-lens survey is CLASS, the Cosmic Lens
All-Sky Survey (e.g.\ Helbig 2000b).  (JVAS is essentially a subset of
CLASS, consisting of the stronger sources.) 
\begin{table}[t]
\caption{Basic statistics of the JVAS and CLASS gravitational-lens
surveys.  `Extra' lens systems are lens systems which were found and
followed up but are not part of the statistically complete sample.  The
`+1' refers to lens candidates which will probably be confirmed.} 
\begin{center}
\begin{tabular}{lccc}
\hline
& JVAS & CLASS & both \\
\hline
sources & 2308 & 6976 & 9284 \\
lens systems in complete sample & 5 & 10+1 & 15+1\\
`extra' lens systems & 1 & 2 & 3\\
total lens systems & 6 & 12+1 & 18+1 \\
\hline
\hline
\end{tabular}
\end{center}
\end{table}
CLASS covers the range in image separation of 0.3--6 arcsec.  It was
later extended up to 15 arcsec, and a smaller survey using a different
strategy extends up to 60 arcsec.  (These additions are not reflected in
the table.)  One wide-separation lens candidate remains (see Phillips et
al., these proceedings). 

While the actual survey is complete, the followup of the lens system is
not.  Thus, here I present only preliminary results: $-0.8 < \lambda_{0}
- \Omega_{0} < 0.3$ (95\%); for $k = 0$: $0.1 < \lambda_{0} < 0.65$.
Note that the {\em lower\/} limit on $\lambda_{0}$ (cf. Quast \& Helbig
1999) in a flat universe is $>$0.  It is important to realise the
difference between constraints being {\em compatible with\/} a certain
cosmological model and {\em favouring\/} a certain cosmological model.
For instance, until recently the constraints in the
$\lambda_{0}$-$\Omega_{0}$ plane from gravitational-lensing statistics
were so broad that the Einstein-de Sitter model was not ruled out.
However, this was never the {\em preferred\/} model; many cosmological
models were compatible with the data.  Also, since the best-fit model
usually occurs in a region of parameter space where the gradient in the
probability density is rather flat, the actual best-fit model is very
sensitive to noise in the input data; more interesting are the much more
stable regions enclosed by a given confidence-level contour. 

\begin{figure}[t]
\plottwo{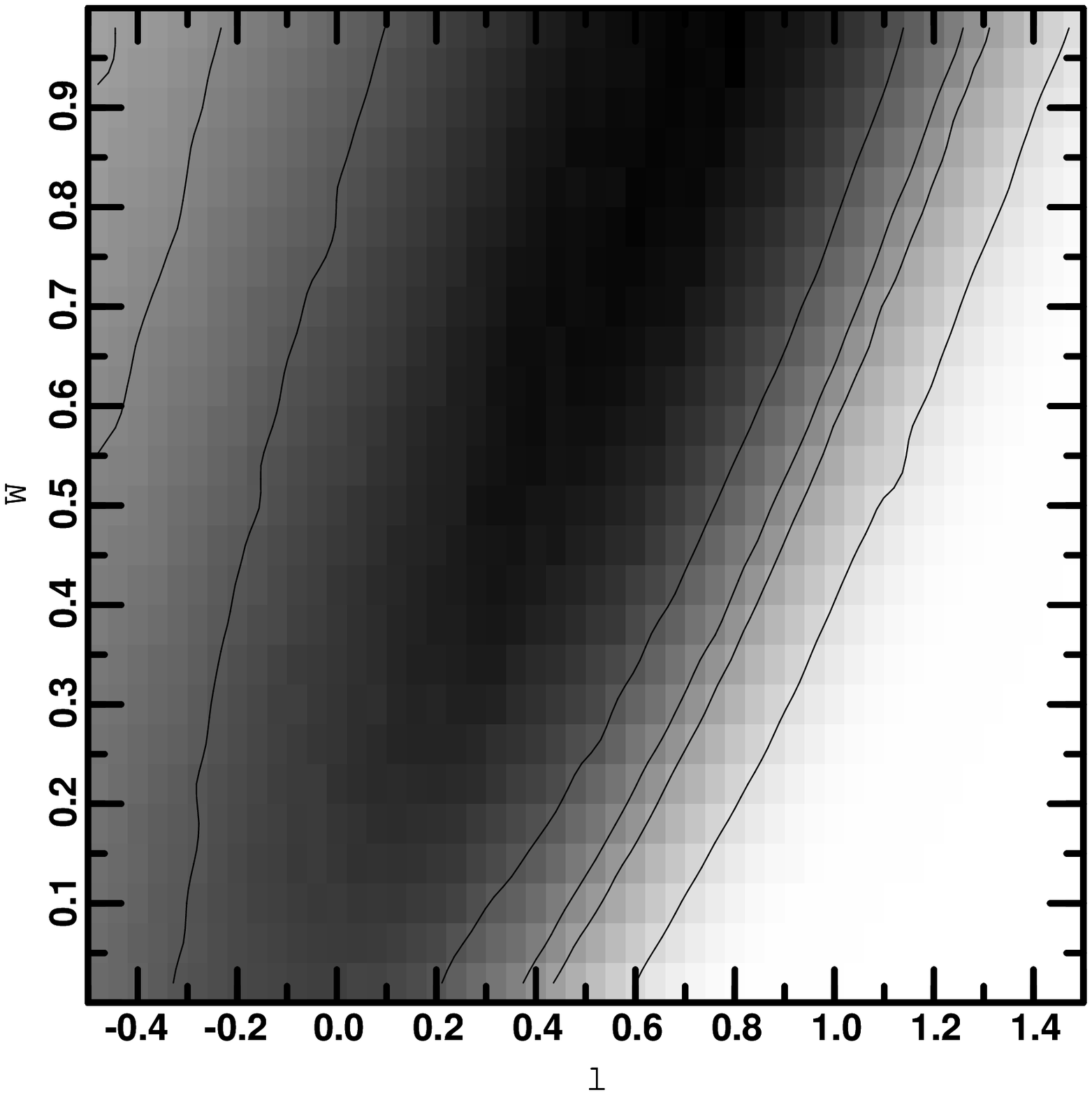}{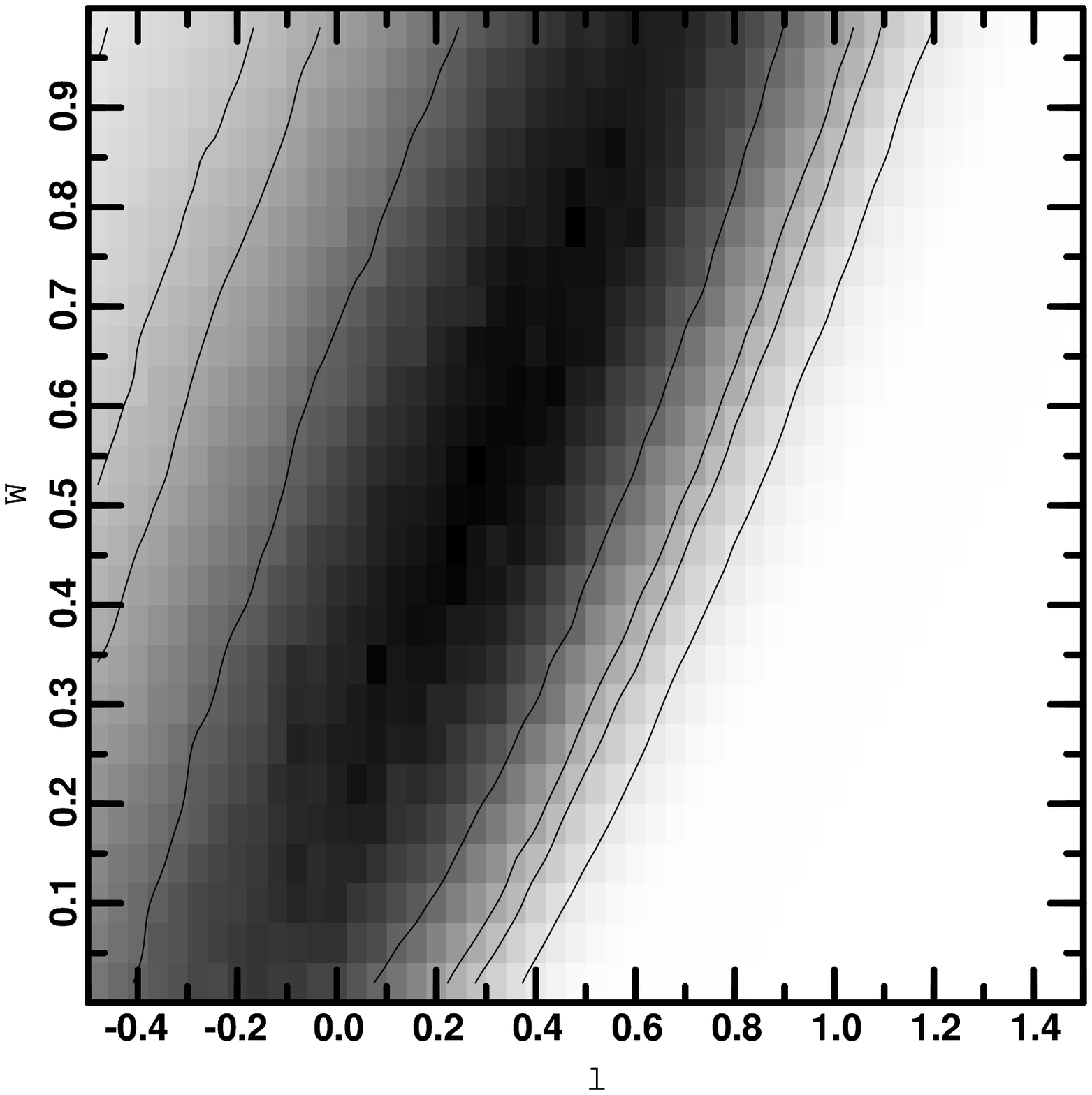}
\caption{Constraints in the $\lambda_{0}$-$\Omega_{0}$ plane from JVAS
(cf. Helbig et al. 1999) (left) and CLASS (excluding JVAS) (right).  
Darker means higher likelihood; contours are at 68\%, 90\%, 95\% and
99\%.} 
\end{figure} 

\begin{figure}[t]
\plotone{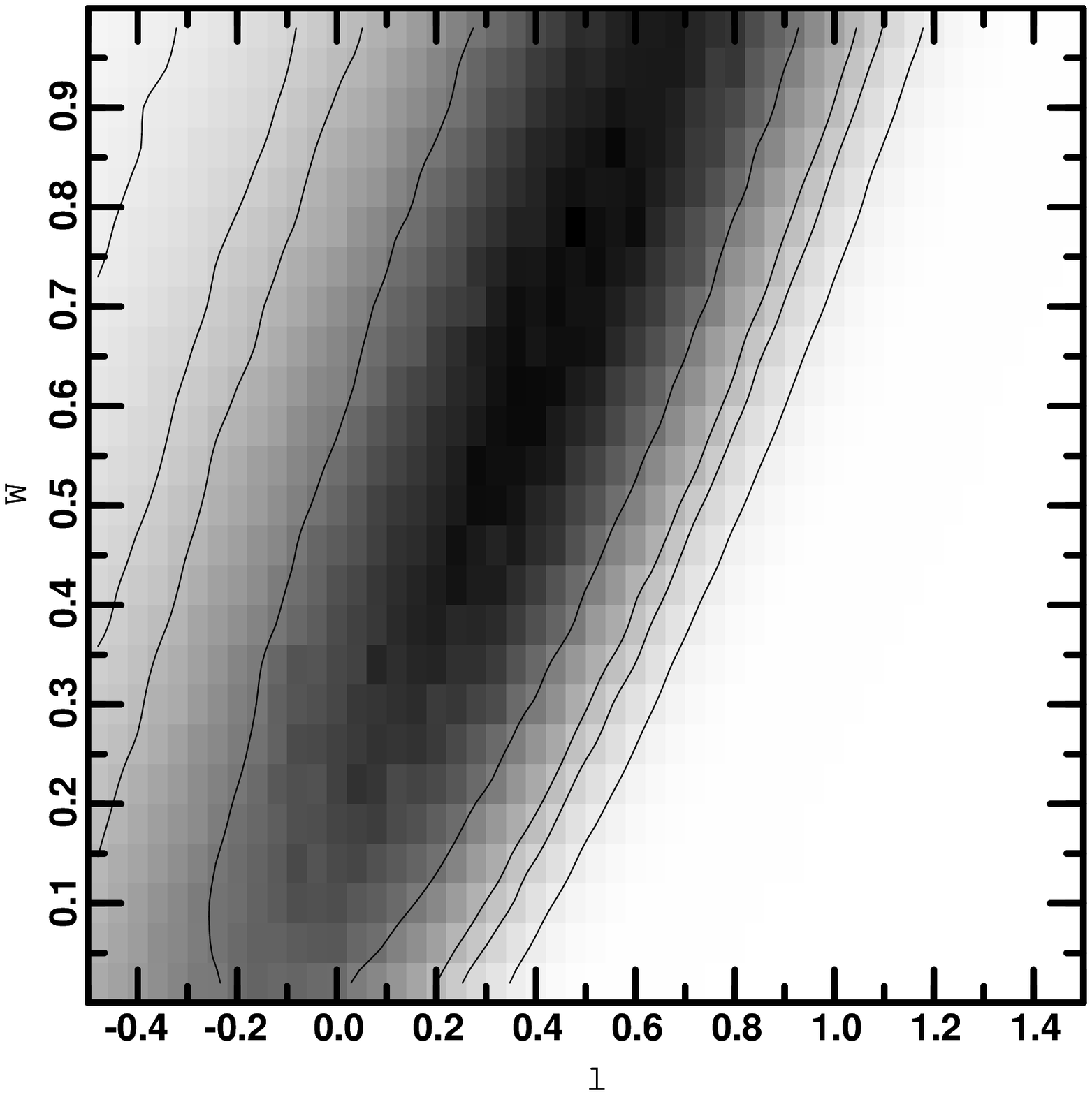}
\caption{Constraints in the $\lambda_{0}$-$\Omega_{0}$ plane using the
entire CLASS (including JVAS).  Darker means higher likelihood; contours
are at 68\%, 90\%, 95\% and 99\%.} 
\end{figure}

A more definitive analysis will be done when all lenses are confirmed
and enough observational data on lens systems and on the source
population are available, taking all errors into account: 
\begin{itemize}
\item statistical errors (including propagation from errors on input
quantities) 
\item systematic errors $\longrightarrow$ statistical errors, i.e.\ we
hope to eliminate systematic errors, mainly due to our ignorance about
details of the source population, by turning them into statistical
errors by observationally constraining the corresponding quantities 
\item sample variance (due to relatively small number of lens systems)
\end{itemize}

\section{Conclusions and Outlook}

\begin{itemize}
\item Despite vast improvements on both the observational and
theoretical sides, the upper limit on $\lambda_{0}$ has been a
remarkably stable number in the astronomical literature. 
\item We can already place a {\em statistically\/} strong upper limit on
$\lambda_{0}$. 
\item Gravitational-lensing statistics is already hinting at a positive
lower limit on $\lambda_{0}$, at least in a flat universe.  While there
is abundant evidence for a positive cosmological constant based on
various combinations of cosmological tests, at present only the $m$-$z$
relation for type Ia supernovae (see e.g. the contributions by
Perlmutter and Kirschner in these proceedings) is the only cosmological
test which indicates this {\em by itself}; it would thus be nice to have
another test which can do so. 
\item When the analysis of the CLASS gravitational-lens survey is
complete, this will improve in the sense that the statistical errors
will become smaller and systematic effects will be reduced; whether or
not the upper limit on $\lambda_{0}$ decreases depends, of course, on
what the value of $\lambda_{0}$ actually is and to what extent current
estimates are biased. 
\item We will have a {\em robust\/} upper limit on $\lambda_{0}$\ soon
when the $S$-$z$ plane for CLASS is better understood.  The $S$-$z$
plane enters into the calculation in two respects, since it determines
both the flux-density dependent redshift distribution, which is needed
to estimate the redshifts of the non-lensed sources in CLASS, and the
redshift-dependent luminosity function, needed for the calculation of
the amplification bias.  (See McKean et al., these proceedings, for some
information about current work in this area.) 
\item We will be able to provide an {\em independent\/} check on
constraints from $m$-$z$ relation (SN Ia), CMB, (evolution of) LSS and
clusters, weak lensing, cluster lensing etc. 
\item Looking farther ahead, many more lens systems will be found in the
future in both radio and optical surveys.  Since the Poisson noise due
to the small number of systems is at present a source of appreciable
uncertainty, progress can be expected just from finding more lens
systems (after doing the corresponding analysis, of course). 
\item Even when the cosmological parameters are known, perhaps more
precisely through (a combination of) other methods,
gravitational-lensing statistics will still be interesting:  It will be
possible to use gravitational-lensing statistics to constrain variables
other than $\lambda_{0}$ and $\Omega_{0}$, and thus study galaxy
evolution, the source population etc. 
\end{itemize}

\acknowledgements

I thank the organisers of IAU Symposium 201 for inviting me to give this
review and the JVAS and CLASS teams for doing the work to provide the
numbers on which my preliminary analysis of the CLASS results is based.
This research was supported by the European Commission, TMR Programme,
Research Network Contract ERBFMRXCT96-0034 ``CERES''.

\end{document}